\documentclass[10pt,conference]{IEEEtran}
\usepackage{cite}
\usepackage{amsmath,amssymb,amsfonts}
\usepackage{algorithmic}
\usepackage{graphicx}
\usepackage{textcomp}
\usepackage{xcolor}
\usepackage{hyperref}
\def\BibTeX{{\rm B\kern-.05em{\sc i\kern-.025em b}\kern-.08em
    T\kern-.1667em\lower.7ex\hbox{E}\kern-.125emX}}
\begin{document}

\title{The Dynamics of Software Composition Analysis}

\author{\IEEEauthorblockN{Darius Foo}
\IEEEauthorblockA{Veracode}
\textit{dfoo@veracode.com}
\and
\IEEEauthorblockN{Jason Yeo}
\IEEEauthorblockA{Veracode}
\textit{jyeo@veracode.com}
\and
\IEEEauthorblockN{Xiao Hao}
\IEEEauthorblockA{Veracode}
\textit{haxiao@veracode.com}
\and
\IEEEauthorblockN{Asankhaya Sharma}
\IEEEauthorblockA{Veracode}
\textit{asharma@veracode.com}}

\maketitle

\begin{abstract}
Developers today use significant amounts of open source code, surfacing the need for ways to automatically audit and upgrade library dependencies, and giving rise to the subfield of Software Composition Analysis (SCA). SCA products are concerned with three tasks: discovering dependencies, checking the reachability of vulnerable code for false positive elimination, and automated remediation. The latter two tasks rely on call graphs of application and library code to check whether vulnerability-specific sinks identified in libraries are used by applications. However, statically-constructed call graphs introduce both false positives and false negatives on real-world projects. In this paper, we develop a novel, modular means of combining call graphs derived from both static and dynamic analysis to improve the performance of false positive elimination. Our experiments indicate significant performance improvements.
\end{abstract}


\section{Motivation}\label{motivation}


Developers today use large amounts of third-party code to build applications, as code reuse significantly increases productivity and lowers development costs \cite{haefliger2008code}. However, given the fact that up to 80\% of typical applications are now third-party code, there is a growing need for tools to manage open source risk \cite{stuart}. The 2017 Equifax data breach was infamously caused by a vulnerability in Apache Struts 2,\footnote{\url{https://nvd.nist.gov/vuln/detail/CVE-2017-5638}} an open source web framework. ``Using Components with Known Vulnerabilities'' is listed in the OWASP Top 10 \cite{owasp}, and the dependency upgrades required to remediate these are time-consuming and often not carried out frequently enough \cite{update-deps} \cite{lauinger2018thou} \cite{auto-pr}. Furthermore, application dependencies constantly change and are difficult to determine and audit manually due to the complexity of modern package managers and library ecosystems.



\emph{Software Composition Analysis} (SCA) is an emerging subfield of application security concerned with precisely this problem. SCA products offer a suite of services centered around automated identification of third-party library dependencies. Auxiliary services such as interfaces for viewing software inventories, enforcing organization-wide policies, and integration with CI/CD \cite{cicd} setups may also be present.

Our SCA product solves two pain points with typical SCA offerings. The first is false positives arising from straightforward dependency analysis -- framework-based applications pull in large trees of transitive dependencies, which, despite the fact that they are included, may not be used in the final application, or at least not in vulnerable ways. We analyze application call graphs to identify and eliminate such cases. The second pain point is remediation: we provide a way to automatically upgrade dependencies, using call graphs again to check if the upgrade is potentially breaking.


In this paper, we describe the architecture of a state-of-the-art SCA product and discuss techniques for performing the core SCA tasks: detection of libraries, reachability of vulnerable code for false positive elimination, and automated remediation. We motivate the need for dynamic analysis to obtain accurate results across all tasks and illustrate a novel means of composing call graphs derived from static analysis and instrumentation in a manner that is modular in third-party libraries, allowing analysis to be performed scalably in CI/CD pipelines. Merging static and dynamic call graphs can result in significantly more vulnerable methods discovered, improving the performance of downstream tasks.







\section{Related Work}\label{related}

The problem of false positives in static call graph construction has been thoroughly investigated \cite{rountev2004static} \cite{frank}, and attempting to improve precision using dynamic analysis is a natural next step \cite{lhotak} \cite{rohit}. Blended analysis \cite{dufour2007blended} \cite{dufour2008scalable} is a similar approach, first obtaining the structure of a program using dynamic analysis, then applying a static analysis on top of it.

\cite{ponta2018beyond} is perhaps the most related work, similarly utilizing a combination of static and dynamic analysis to eliminate false positives in SCA. The key innovations of our approach are the use of hand-curated, vulnerability-specific sinks and the fact that the analysis is modular, not requiring call graph construction on demand for libraries (an idea similar to \cite{modular}). The latter makes the analysis scalable, allowing it to complete in minutes on average and making running it in a CI pipeline feasible.

\section{Discovering Dependencies}\label{deps}

An essential problem in SCA is that of \emph{discovering dependencies}: determining the third-party libraries a project uses given its source code. Results are typically drawn from some universe of open source libraries, such as coordinates on Maven Central.

\subsection{Static Dependency Analysis}

The most straightforward way of determining a project's dependencies is to read its \emph{dependency manifests}, e.g.~\texttt{pom.xml}. These contain a listing of library coordinates and associated version constraints. Package managers interpret these manifests to perform \emph{dependency resolution}: querying an external repository to determine the transitive dependencies of each library (which may themselves introduce more constraints), then selecting a set of libraries which satisfies all constraints. In the event of an inability to satisfy every constraint, package managers may fail or approximate a solution (e.g.~Maven's \emph{nearest definition} heuristic, which applies when multiple versions of the same library are required). The definition of what constitutes a valid solution also varies, e.g.~npm does not require a single version of each library.

\subsection{Dynamic Dependency Analysis}

The main difficulty of detecting dependencies statically is in modeling the implementation of a package manager correctly. Package managers are typically unspecified and must be each handled specially. Dependency resolution is also often nondeterministic and might produce different results over time, due to reliance on external repositories that may be updated by developers. This suggests that a better approach is to perform a dynamic analysis instead; a theme we explore in this paper.

A dynamic dependency analysis integrates with package managers and can get the results of dependency resolution exactly as they would have produced at a point in time. The main shortcoming is that a full build of the project may be required, which is time-consuming and difficult to automate in general, due to the system dependencies and configuration required.



Nevertheless, going from a purely static analysis to a package-manager-integrated dynamic analysis demonstrably improves accuracy. We evaluate this using a set of 41 microbenchmarks\footnote{\url{https://github.com/srcclr/efda}} across 18 different package managers. On average, we see that the dynamic analysis improves the number of dependencies discovered by 204\% (131 with purely static, and 398 with dynamic), with no cases in which it performs worse.


\section{Identifying Vulnerabilities}\label{vulns}

\subsection{Vulnerability Databases}

The next step is to identify vulnerabilities due to libraries. A baseline data source is the NVD; practically every SCA product informs users of CVE vulnerabilities. SCA vendors also curate their own vulnerability datasets, supported by both data mining methods \cite{ml1} \cite{sabetta2018practical} and custom tooling \cite{sgl}.

Determining if a vulnerability is exploitable in an application would allow an SCA product to report fewer false positives. This is valuable because notification fatigue from false positives becomes the bottleneck when fixes are deployed automatically \cite{auto-pr}. Furthermore, the amount of false positive reduction is significant: commercial SCA products report that 70-80\% of dependencies are never referenced in application code.



Our approach is based on a lightweight reachability analysis using call graphs. We discuss the tradeoffs of various call graph construction methods before covering our implementation of the analysis and how it fits into our SCA product.

\subsection{Static Call Graphs}

Given an application's source code, we may view it as a \emph{call graph}, a structure that expresses the calling relationships between its methods. A directed edge $e$ from method $m_1$ to method $m_2$ indicates that a call site within $m_1$ invokes $m_2$ in some execution of the program. A \emph{call chain} is a sequence of edges $e_1 e_2 ... e_n$ where $(m_n, m_{n+1}) = e_n$ and $(m_{n+1}, m_{n+2}) = e_{n+1}$, i.e. consecutive edges must share a method vertex.

Call graph construction for object-oriented programs must model the effects of dynamic dispatch correctly. False positives are possible when applying such analyses \cite{grove2001framework}, leading to \emph{infeasible} edges which are present in the graph but do not occur in any concrete run of the program. Furthermore, \emph{call chains} comprising only feasible edges may themselves prove be infeasible. On average, 25.5\% of call chains created by static approaches are infeasible \cite{rountev2004static}.

The presence of infeasible call chains means that a reachability analysis would produce false positives, traversing paths which will never occur at runtime.

\subsection{Dynamic Call Graphs}


Real-world static analysis tools often deliberately eschew prohibitively expensive or conservative approximations that would render analysis less useful when ``hard'' language features such as reflection are involved \cite{livshits2015defense}. This causes problems when analyzing highly dynamic frameworks, such as Spring or Rails: the resulting call graphs are incomplete and lead to false negatives in reachability analyses.

An alternative is the use of runtime instrumentation, a form of dynamic analysis. Executing a program's tests and recording its flows has the upside of never producing any infeasible call graph edges, since only flows which were actually observed are reported. Support for all language features is also a given.

The downsides are that this approach produces false negatives (missing paths to sinks called by code not covered by tests) and requires code to actually build and run in order to be analyzed. Instrumentation also imposes overhead, slowing test execution.


\subsection{Combined Static and Dynamic Graphs}

\begin{figure}
\centering
\includegraphics[scale=0.3]{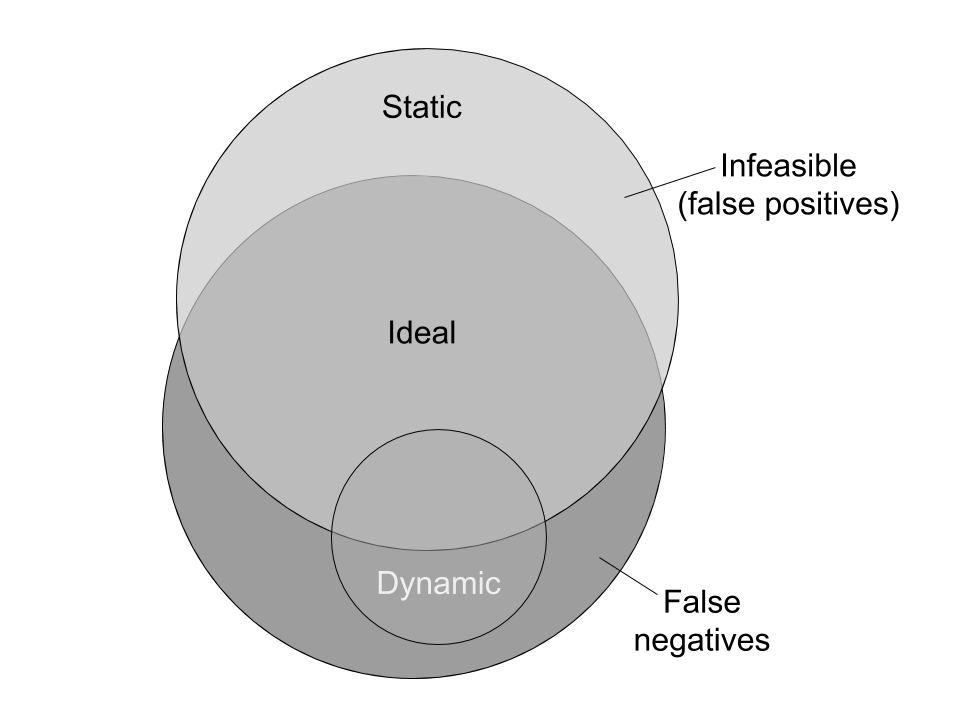}
\caption{Combined static and dynamic call graphs}
\label{fig:combined}
\end{figure}

The relationship between static and dynamic call graphs, as well as how they approximate a \emph{theoretically ideal} call graph, has been investigated by \cite{lhotak}. In short, the ideal call graph is the union of dynamic call graphs across all possible executions of a program (making the dynamic call graph for one execution a subset of it), and the static call graph is a superset of the ideal call graph due to the presence of infeasible edges. Due to the soundiness \cite{livshits2015defense} of our approach (because of false negatives from unsupported language features; covered in detail in Section~\ref{static-cg}), we modify the definition slightly, so that neither the static call graph nor the ideal call graphs are subsets of each other. The modified relationship is shown in Figure~\ref{fig:combined}.

The final call graph that we search for uses of vulnerability-specific sinks is derived by taking the union of the static and dynamic call graphs, the intuition being that it brings us closer to the ideal call graph. In the following sections, we explain how this is done in a modular fashion, preserving the properties of the graph that allow a fast reachability check.

\subsection{Vulnerable Methods}\label{vuln-methods}

We consider a vulnerable library to be \emph{possibly used} in an application if a vulnerability-specific sink is reachable from the application's call graph. This is weaker than determining if a vulnerability is \emph{exploitable} as control flow is not taken into account, admitting false positives. We curate these sinks by hand; these are the method-level root causes of vulnerabilities mentioned in CVEs and proprietary vulnerabilities.

\subsubsection{Static Call Graph Construction}\label{static-cg}

\begin{figure}
\centering
\includegraphics[scale=0.4]{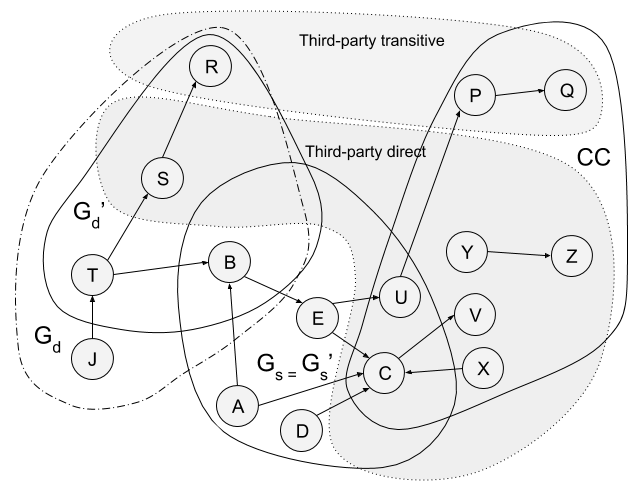}
\caption{Example of a composed call graph. The boundary lines represent (from left to right) the sets $G_d$, $G_d'$, $G_s = G_s'$, methods of a direct dependency, methods of a transitive dependency, and CC.}
\label{fig:composed}
\end{figure}

To illustrate our approach, we refer to Figure~\ref{fig:composed}, which shows the final graph, with labels for vertices (e.g.~A) and contours for subgraphs (e.g.~CC). Given an application, we start by constructing a static call graph $G_s = (V_s, E_s)$, represented by the contour $G_s$. As we only analyze the application, $V_s$ consists of both application methods (B, A, D, E) and the entry points of libraries called \emph{directly} by the application (C, U); transitively-called library methods (V, P, Q) are absent.

$E_s$ is initialized to the set of statically-known static or virtual calls; given a method call $a.b()$ in method $m$, we add an edge between $m$ and the method $b$ of $a$, using $a$'s declared class or interface.

We expand the set $E_s$ with a number of passes:

\begin{enumerate}
  \item Class Hierarchy Analysis (CHA) \cite{cha}, which determines possible receiver classes for $b$ using subclass relations
  \item Rapid Type Analysis (RTA) \cite{vta}, which rules out receiver classes using information about object instantiations
  \item Reflection analysis, which adds edges for reflective calls with constant arguments; as this is a subset of all possible edges due to reflective calls, the approach is soundy \cite{livshits2015defense}
\end{enumerate}


Thus we have $G_s' = (V_s, RTA(CHA(E_s)))$. We define the set of \emph{first-party entry points} $V_e$ as the set of methods without callers, i.e.~$\left\lbrace m_1 \mid \exists m_1 \in Vs, \forall (m_2, m_3) \in E_s, m_1 \neq m_3 \right\rbrace$. A and D in the diagram are examples. Methods of $V_e$ must be first-party, as third-party methods must be called by a first-party method to appear at all. $V_e$ may be further filtered down (e.g.~by considering only \texttt{main} methods) depending on analysis goals. $G_s'$ is constructively a graph of all methods reachable from a first-party entry point.

Separately, we precompute \emph{vulnerable method call chains} $CC$ for libraries: sequences of directed edges $e_1 e_2 ... e_n$ such that $e_n$ ends at a vulnerable method (Z, V, Q), and $e_1$ is an entry point of the library (Y, X, U, P). A call chain represents a path from a library entry point to a library-specific sink.

\subsubsection{Merging Library Call Chains}

Given a vulnerable method call chain $e_1 e_2 ... e_n$ and a static call graph $G_s'$, we merge them using the algorithm in Figure~\ref{fig:callchain}. For each call chain, we iterate through its edges (line 1), looking for a suffix which begins at an existing edge in the graph (lines 2-3), adding the suffix to the graph if we find it (line 4).

\begin{figure}
\centering
\begin{minipage}{.7\linewidth}
\begin{algorithmic}[1]
\FOR{$e$ in $e_1 e_2... e_n$}
\STATE{$(m_1, m_2) \leftarrow e$}
\IF{$m_1 \in V_s$}
\STATE{add $e ... e_n$ to $G_s'$}
\STATE{\textbf{break}}
\ENDIF
\ENDFOR
\end{algorithmic}
\caption{Merging vulnerable method call chains}
\label{fig:callchain}
\end{minipage}
\end{figure}

Applied to $G_s'$, this would result in V being added. The approach ensures that we do not introduce third-party entry points (X, Y, P): new outgoing edges are only attached to existing ones and no new incoming edges are added. This preserves the property that all vertices are reachable from a first-party entry point. Consequently, determining if a vulnerable method is called in some execution of the program -- the key question we are interested in answering -- is a simple membership check.


\subsubsection{Dynamic Call Graph Construction}

Given tests, we instrument them to derive a dynamic call graph, then compose it with the static call graph and vulnerable method call chains to find more potentially reachable vulnerable methods.

Running the tests, we construct a dynamic call graph $G_d = (V_d, E_d)$. $V_d$ comprises both first- and third-party methods (J, T, S, R, B), and $E_d$ contains only feasible edges and paths. In contrast to $E_s$, where method callers were only ever first-party methods and callees were either first- or third-, $E_d$ contains third-party callers \emph{and} callees -- this may be observed from the execution of a framework such as JUnit, where a \texttt{main} method defined in JUnit itself (J) is called, which dynamically discovers \texttt{@Test}-annotated user methods (T) to invoke reflectively. This \emph{inversion of control} is a common pattern in framework-based applications \cite{ponta2018beyond}. First-party entry points (T) thus \emph{do} have callers.


When composing the static and dynamic call graphs, we would like to preserve the property that allowed us to determine reachability using set membership. We have to add entry points now, though, because tests were never considered earlier; the solution is then to ensure that they are first-party.

We first instrument tests, and when the test run completes (either in success or failure), the resulting edges form the graph $G_d$. Next we identify \emph{framework entry points}: these are the edges which span framework and application code (JT). We identify frameworks manually here as there is no good way to differentiate them without more context, special-casing common ones like JUnit and TestNG.

Finally, given a framework entry point, we take the vertices of its transitive closure (T, S, R, B) and add it to a new graph $G_d'$ -- a graph of dynamic edges whose entry points are first-party.

The union of $G_s'$ and $G_d'$ gives us $G_c$, a combined static and dynamic call graph with only first-party entry points. We may further merge the static call chains into $G_c$. Paths in $G_c$ may then span both static and dynamic edges.



\subsection{Discussion}

A limitation of vulnerable method call chains being computed for single libraries in isolation is that vulnerable methods called \emph{only within third-party code} will be missed. For example, the edge UP will not be in any call chain because we cannot tell when computing call chains that P is the start of a vulnerable method call chain from another library. The dynamic call graph does handle this case, however, as shown by S and R; if there were an edge RP (analogous to UP), we would be able to detect the vulnerable method call.











\section{Remediation}\label{remediation}

Another task we perform automatically is remediate library vulnerabilities. Our approach \cite{update-advisor} is a static analysis that precomputes diffs between library versions to determine the changes between them. The diffs are augmented with call graphs (constructed as in Section~\ref{vuln-methods}) and are thus \emph{semantic} in nature, e.g.~considering methods changed if their callees change.


We then check if methods in the diffs of potential library upgrades occur in the application's call graph. These are shown as potential breaking changes in pull requests that we create automatically, indicating to developers which upgrades are riskier.

\subsection{Dynamic Analysis}

In keeping with the theme of the paper, we outline an extension to the above analysis which uses an instrumentation-derived call graph to improve accuracy. The analysis depends on accurate call graphs when computing diffs with semantic information, and checking if an application uses a method of a combined diff. The latter may benefit from this, especially if there are tests available. Given that a dynamic call graph will not contain infeasible edges, there will be not be additional false positive calls to libraries, leading to fewer library upgrades being considered as breaking when they are not.


There is also the obvious ``dynamic analysis'': executing tests to check if an upgrade introduces breakage. This does not subsume the entire analysis for breaking changes as the latter reveals changes occurring outside the coverage of the test suite and non-breaking semantic changes.

\section{Evaluation}\label{eval}

\begin{table}[!htbp]
\caption{Call graph edges}
\centering
\begin{tabular}{|l|l|l|l|l|l|l|}
\hline
\textbf{project} & \textbf{\shortstack{static \\ vertices}} & \textbf{\shortstack{static \\ edges}} & \textbf{\shortstack{dynamic \\ vertices}} & \textbf{\shortstack{dynamic \\ edges}} & \textbf{\shortstack{static \\ sinks}} & \textbf{\shortstack{dynamic \\ sinks}} \\ \hline
helios           & 7930                     & 26746                 & 12287                     & 39813                  & 1                     & 3616                   \\ \hline
immutables       & 30206                    & 319934                & 398                       & 874                    & 5                     & 5                      \\ \hline
java-apns        & 536                      & 999                   & 4240                      & 8685                   & 58                    & 859                    \\ \hline
retrofit         & 1925                     & 5269                  & 7339                      & 22565                  & 6                     & 6                      \\ \hline
\end{tabular}
\label{table:callgraph}
\end{table}

We evaluated the performance of call graph construction on four real-world Maven-based Java projects (Table~\ref{table:callgraph}). On average, we find that dynamic call graphs add 824\% more vertices and 361\% more edges, allowing us to discover significantly more call sites from which vulnerable methods are reachable in 2/4 cases (shown in the \textbf{static sinks} and \textbf{dynamic sinks} columns). Most of the extra edges are from third-party dependencies. The tradeoff is that dynamic call graphs less easy to apply automatically, as manual effort is required to configure projects so that they compile successfully and their tests run at least partially. There is also significant variance between projects in test coverage (and correspondingly, dynamic call graph size and vulnerable method reachability).

\section{Conclusion and Future Work}\label{conclusion}

We motivated and described the SCA problem: the fact that major portions of modern applications are third-party emphasizes the need for tooling to automatically audit and upgrade dependencies. Our approach has three components: dependency analysis, call-graph-based analysis of library use augmented with hand-annotated library-specific sinks, and automated remediation. We illustrate the need for dynamic analysis in each of these tasks, using package manager integrations to identify dependencies and instrumentation to build dynamic call graphs. We also describe a novel means of composing the dynamic and static call graphs together with precomputed call chains, making the analysis modular in the library dependencies of the application. This significantly improves the ability to find vulnerable methods.

In future, we hope to focus on automated remediation, for example by performing transitive dependency upgrades, or optimizing upgrades by pruning redundant dependencies or reacting to dependency conflicts \cite{dependency-conflicts}.

\vspace{12pt}

\end{document}